\documentclass[nocompress]{spie}  
\usepackage{xcolor}
 
\usepackage{amsmath}  
\usepackage{textcomp} 
\usepackage{graphicx}
\usepackage[colorlinks=true, allcolors=blue]{hyperref}

\title{ESCAPE project: testing active observing strategies for high-contrast imaging in space on the HiCAT testbed}
\author[a]{Alexis Lau}
\author[a]{Élodie Choquet}
\author[a]{Lisa Altinier}
\author[b]{Iva Laginja}
\author[c]{Rémi Soummer}
\author[c]{Laurent Pueyo}
\author[a]{Nicolas Godoy}
\author[a]{Arthur Vigan}
\author[d]{David Mary}

\affil[a]{Aix Marseille Univ, CNRS, CNES, LAM, Marseille, France}
\affil[b]{LESIA, Observatoire de Paris, Universit\'{e} PSL, Sorbonne Universit\'{e}, Universit\'{e} Paris Cit\'{e}, CNRS, 5 place Jules Janssen, 92195 Meudon, France}
\affil[c]{Space Telescope Science Institute, 3700 San Martin Drive, Baltimore, USA}
\affil[d]{Université Côte d’Azur, Observatoire de la Côte d’Azur, CNRS, Laboratoire Lagrange.}

\authorinfo{Further author information, send correspondence to A.L.: alexis.lau@lam.fr}

\begin{document} 
\maketitle

\begin{abstract}
The Roman Space Telescope will be a critical mission to demonstrate high-contrast imaging technologies allowing for the characterisation of exoplanets in reflected light. It will demonstrate $10^{-7}$ contrast limits or better at 3--9 $\lambda / D$ separations with active wavefront control for the first time in space. The detection limits for the Coronagraph Instrument are expected to be set by wavefront variations between the science target and the reference star observations. We are investigating methods to use the deformablel mirrors to methodically probe the impact of such variations on the coronagraphic PSF, generating a PSF library during observations of the reference star to optimise the starlight subtraction at post-processing. We are collaborating with STScI to test and validate these methods in lab using the HiCAT tested, a high-contrast imaging lab platform dedicated to system-level developments for future space missions. In this paper, we will present the first applications of these methods on HiCAT.
\end{abstract}


\section{INTRODUCTION}
\label{sec:intro}  
Thousands of exoplanets have been detected through indirect methods such as radial velocity and transit measurements. However, studies aiming to detect and characterise Earth analogues around main-sequence stars will require direct imaging observations, which face significant challenges due to the small angular separation between the host star and the exoplanet, and the high planet-to-star flux ratio. Although coronagraphs and high-order wavefront sensing and control (HOWFSC) have been demonstrated under laboratory conditions and in ground-based observatories, their performance is primarily limited by absorption features in certain band-passes and atmospheric turbulence.

The James Webb Space Telescope (JWST) undoubtedly has a significant advantage in wavelength coverage from $\lambda$ \textgreater 3 to the mid-infrared range, owing to its stability and space-based environment. Work led by Carter \cite{Carter2023} demonstrates how JWST enables deeper atmospheric characterisation with the Mid-Infrared Instrument (MIRI) and Near-Infrared Camera (NIRCam), although JWST's high-contrast capabilities could be enhanced by reducing residual starlight at close separations \cite{GodoySPIE2024}.

Post-processing algorithms, such as KLIP \cite{Soummer2012} aim to completely remove such residuals and bring the detection down to the photon noise level at close separations. In addition to post-processing algorithms, there are observing strategies to introduce image diversities for high-contrast image post-processing, including Angular Differential Imaging (ADI) \cite{Marois2006}, Reference Differential Imaging (RDI) \cite{Smith1984}, and Coherent Differential Imaging (CDI) \cite{Potier2022, Bottom2017, Baudoz2006}. Most of these methods were developed for ground-based telescopes and utilise image diversity to build empirical point spread function (PSF) models. With post-processing, the current detection limits are typically at $\approx$ $10^{-6}$ for JWST and ground-based imagers.

The launch of the Nancy Roman Space Telescope will be a key milestone in direct imaging, as it will be the first equipped with an advanced coronagraph, featuring deformable mirrors to perform wavefront control and sensing.

The Roman Coronagraph is expected to be capable of detecting reflected visible light from mature, giant exoplanets and performing characterisation, owing to the novel implementations of wavefront control in space. Acting as a technological demonstration, the Roman Coronagraph is anticipated to bridge the gap between current high-contrast imaging capabilities and those expected for future missions such as the Habitable World Observatory (HWO), recommended by the decadal survey to image Earth-like planets. The Roman Coronagraph is expected to achieve a contrast level of $10^{-7}$ at 6 - 9 $\lambda / D$ (minimum requirement \cite{Krist2023}), with HOWFS and control algorithms, including Electric Field Conjugation (EFC) \cite{Give'on2007SPIE} combined with Pair-wise Probing \cite{Giveon2011SPIE}, to attain the desired level of contrast. However, the contrast attained is expected to degrade when HOWFSC is not active during the observing phase, while low-order wavefront sensing and control operates at 2 mHz.

The baseline post-processing method to remove the residual starlight in the dark zone and mitigate the effects of wavefront errors (WFEs) is to use classical RDI or RDI-PCA subtraction to disentangle astrophysical signals from quasi-static speckles. The Roman Coronagraph presents an ideal opportunity for demonstrating novel observing strategies and post-processing techniques that incorporate telemetry data in high-contrast imaging. The objective of this project is to develop calibration methods using Deformable Mirrors (DMs) and examine their effects on coronagraphic Point Spread Functions (PSFs). In this paper, we discuss the upcoming implementation and demonstration of such post-processing techniques in a lab environment. 

\section{Implementation on the HiCAT testbed}
\label{sec:implement-hicat}
\begin{figure}[!htt]
    \centering
    \includegraphics[width=.8\textwidth]{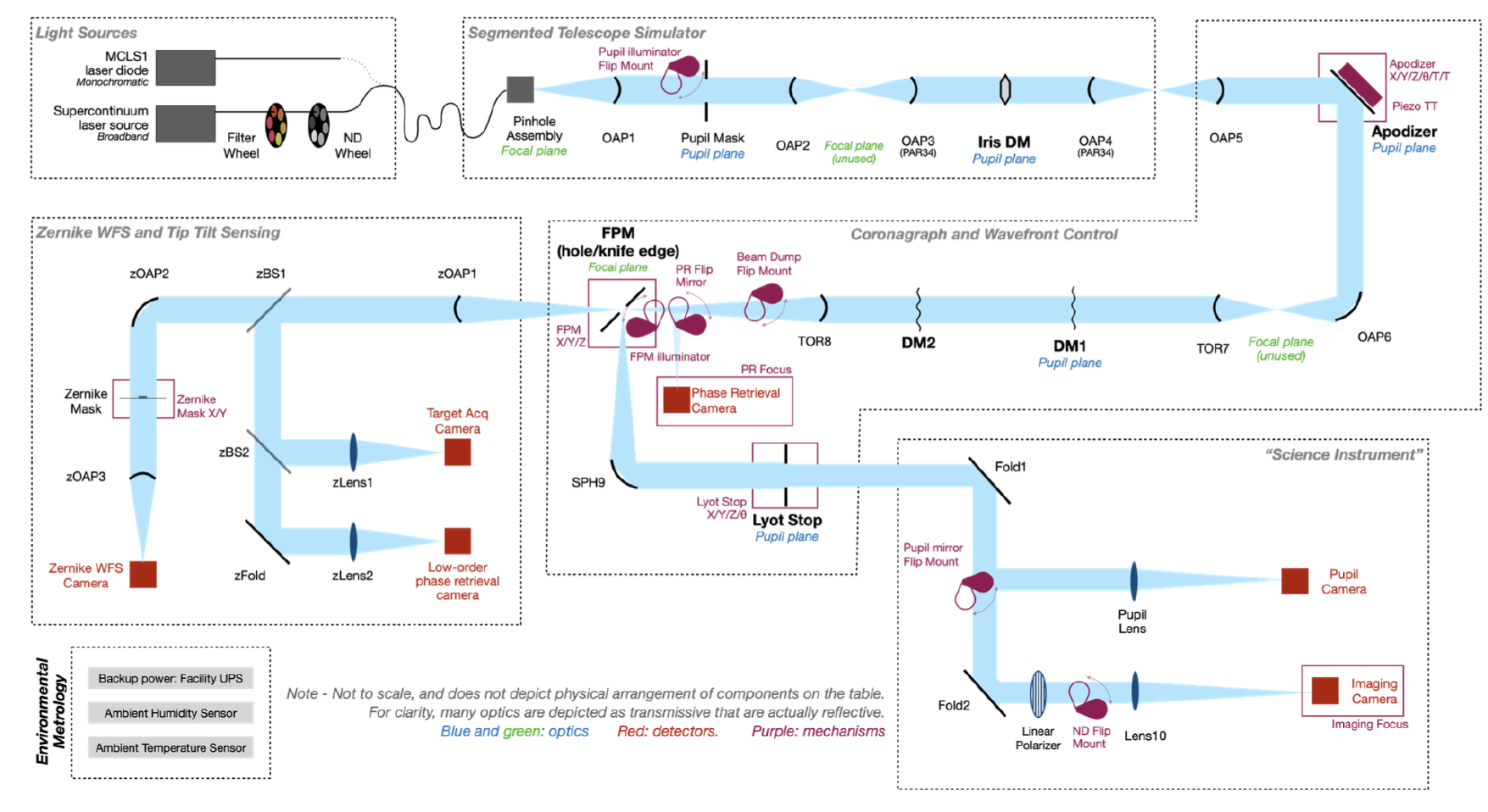}    
    \caption{The HiCAT layout reported in the SPIE proceeding in 2022 by the HiCAT team \cite{SoummerSPIE2022}.}
    \label{fig:hicatscheme}
\end{figure}

Our focus is on demonstrating coronagraphic PSF calibrations with DM probes for space-based telescopes. Given that such a facility does not exist, optical benches like the HiCAT (High Contrast Imager for Complex Aperture Telescopes) serve an important role when it comes to technological demonstrations and validations.Figure \ref{fig:hicatscheme} shows a schematic of the HiCAT testbed.

\subsection{Brief Overview of HiCAT}
The HiCAT testbed is a specialised facility featuring on-axis segmented apertures designed for system-level high-contrast imaging experiments under ambient conditions. It is equipped with a Zernike Wavefront Sensor (WFS), which enables it to maintain dark hole quality while compensating for low-order drifts. For HOWFC, focal-plane wavefront sensing is implemented, using the pairwise probing and stroke minimisation algorithms. Mid-order WFC was implemented this year, detailed in \cite{SoummerSPIE2024}. As detailed in the SPIE proceeding in 2024 \cite{SoummerSPIE2024}, the testbed routinely achieves $2.5 \times 10^{-8}$ in narrowband (3 $\%$ at $\lambda$ = 680 nm) with an Apodised-Pupil Lyot Coronagraph at 4.4 - 11 $\lambda / D_{pupil}$, where the typical contrast in broadband (9 $\%$ at $\lambda_{centred}$ = 680 nm) $6 \times 10^{-8}$. These dark holes can be created within a few minutes starting from uncontrolled deformable mirrors (DMs).

The HiCAT testbed is operable and scriptable via Python, complemented by a Graphical User Interface (GUI). The backend and communication layer, CATKit2 \cite{catkit2}, are implemented in C++ to facilitate faster connections between 'services', defined in CATKit2. Besides controlling the testbed, the GUI allows users to monitor the status and output of various services on the bench. Experiments such as broadband Electric Field Conjugation \cite{Give'on2007SPIE} (EFC) can also be initiated through the GUI. The latest update regarding the results and the performance of HiCAT are reported in this SPIE \cite{SoummerSPIE2024}.

In order to prepare demonstrations on the hardware, HiCAT has a digital twin that simulates to a high level of accuracy the actual hardware implementation, including the instrument propagation model, using the Python-based HCIPy\cite{Por2018HighContrastImaging} package. This is used to develop and test experiments before safely deploying them on the testbed. 

\subsection{Designing Experiments to be Tested on HiCAT}
\begin{figure}[!ht]
    \centering
    \includegraphics[width=\textwidth]{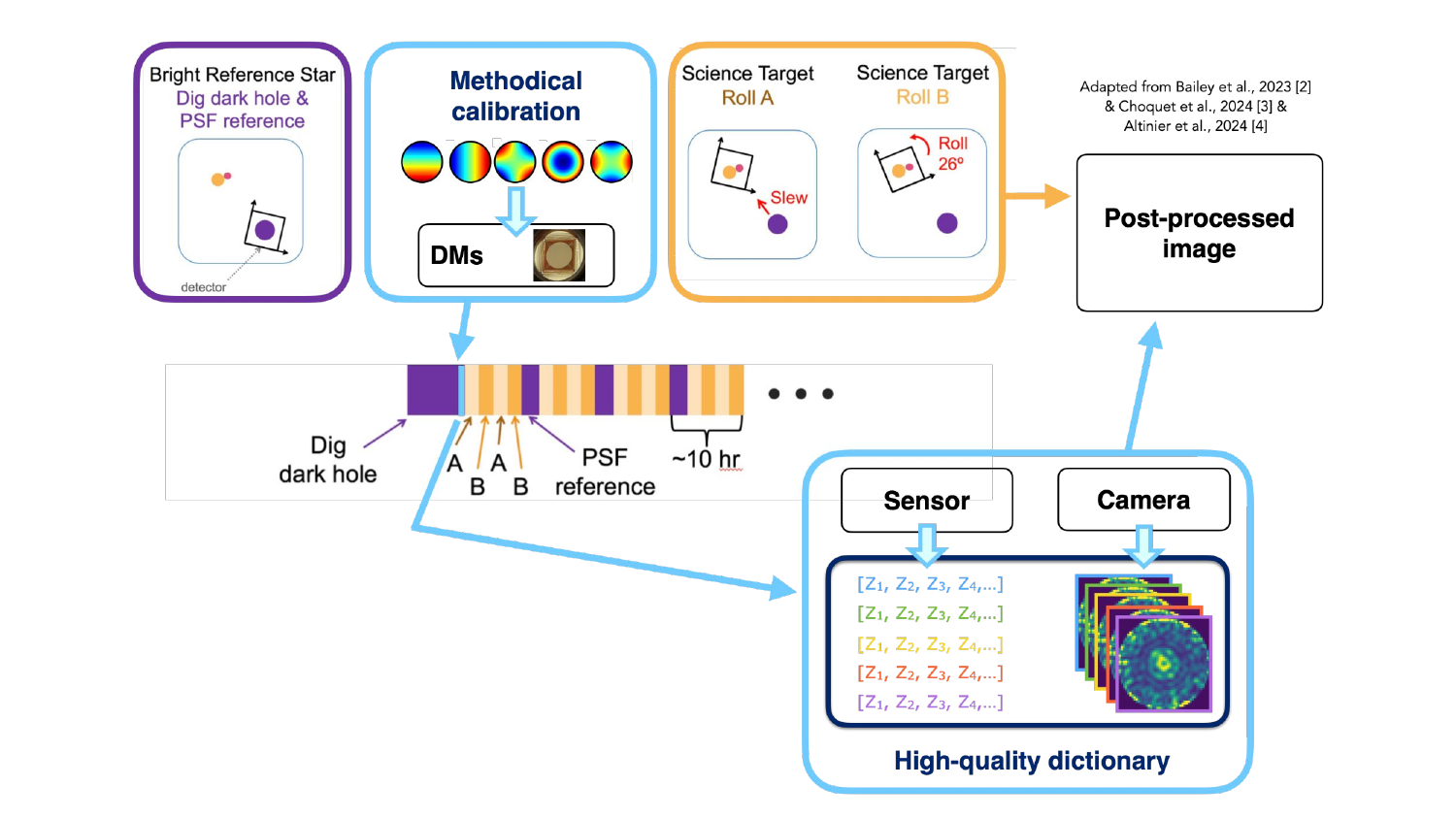} 
    \caption{Methodology of injecting calibration probes during reference acquisition.}
    \label{fig:ESCAPE_method}
\end{figure}

As part of the ESCAPE (Exoplanetary Systems with a Coronagraphic Archive Processing Engine) project, the objective is to develop and demonstrate improvements in starlight subtraction through post-processing methods that utilise hardware in high-contrast imagers. To prepare these methods for high-contrast imaging in space, we first implement the observing sequence on the testbed, examining its performance with active WFS and control (WFS$\&$C) and establishing baseline performance using classical post-processing techniques. By doing so, we assess the impact of the injected modes with the DMs on the coronagraphic Point Spread Functions (PSFs) and determine whether these maps can aid in calibrating the coronagraphic PSFs during post-processing. Figure \ref{fig:ESCAPE_method} illustrates our proposed methodology for enhancing current post-processing techniques.

\begin{figure}[!ht]
    \centering
    \includegraphics[width=1\textwidth]{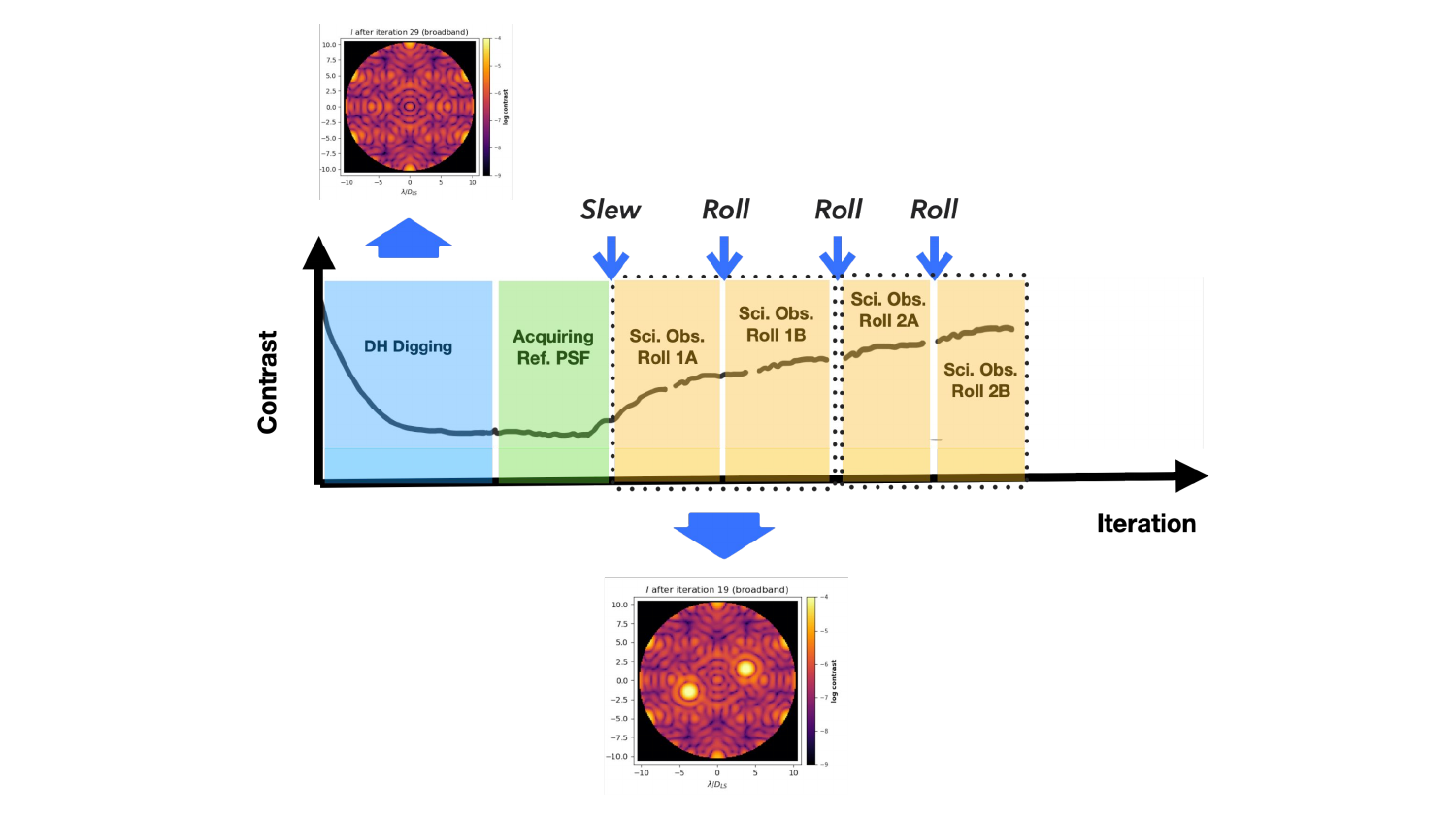}    
    \caption{Schmematic of the contrast evolution during the observing sequence.}
    \label{fig:hicat_experiment}
\end{figure}

To investigate the key components for improving starlight subtraction, we first adapt the observing sequence from the Roman Coronagraph into experiments to be conducted on HiCAT. These experiments are designed using the digital twin of HiCAT, developed by the STScI team. Figure \ref{fig:hicatscheme} presents a schematic of the experimental setup. Following the procedures of the Roman Coronagraph, the experiment is segmented based on specific operations. Hence, a dark hole is created by performing EFC, followed by reference PSF acquisition. Subsequently, additional wavefront errors are added on the DM to simulate the wavefront drift associated with the telescope slew to a science target, and planets are injected into the sequence using DMs to simulate potential detections at varying contrast levels. To replicate the procedures employed on the Roman Coronagraph, additional wavefront errors are also added to simulate the telescope roll done during the science observation to introduce angular diversity.

During the process of digging the dark hole, we save the results from the estimator and the controllers, as well as the shapes and commands of the deformable mirrors, along with the coronagraphic PSFs. This same information is saved during the science acquisition stage.

By studying the response of the coronagraphic point spread functions to the injected modes and acquiring such telemetry data, we aim to use this information to build libraries for calibrating PSFs during post-processing, utilising deformable mirrors for space missions. Thus far, using the HiCAT Digital Twin, we have successfully created the dark hole with the targeted contrast and implemented the sequence with planet injections without additional aberrations and rolling between targets using the numerical simulator. The next steps include introducing a roll angle between each science observation.

\section{Conclusion and Perspectives}

In the coming months, we will introduce additional, controlled aberrations in the numerical simulator, then commence our experiments on the real hardware testbed to investigate the impact of DM probes on the coronagraphic PSFs. Subsequently, we will assess how combinations of probes affect the focal plane image, determine whether we can recover planets when the planetary signal is faint. If possible, we will also incorporate low-order WFS$\&$C.

Alongside these experiments, we will begin developing PSF calibrations within post-processing methods and compare the results with classical post-processing techniques such as PCA and RDI. To apply sensor fusion in post-processing, we will develop telemetry classes for the experiments and start creating libraries using the experimental results.

\acknowledgments 
This project is funded by the European Union (ERC, ESCAPE, project No 101044152). Views and opinions expressed are however those of the authors only and do not necessarily reflect those of the European Union or the European Research Council Executive Agency. Neither the European Union nor the granting authority can be held responsible for them.

\bibliography{report.bib} 

\begin{thebibliography}{10}

\bibitem{Carter2023}
{Carter}, A.~L., {Hinkley}, S., {Kammerer}, J., {Skemer}, A., {Biller}, B.~A., {Leisenring}, J.~M., {Millar-Blanchaer}, M.~A., {Petrus}, S., {Stone}, J.~M., {Ward-Duong}, K., {Wang}, J.~J., {Girard}, J.~H., {Hines}, D.~C., {Perrin}, M.~D., {Pueyo}, L., {Balmer}, W.~O., {Bonavita}, M., {Bonnefoy}, M., {Chauvin}, G., {Choquet}, E., {Christiaens}, V., {Danielski}, C., {Kennedy}, G.~M., {Matthews}, E.~C., {Miles}, B.~E., {Patapis}, P., {Ray}, S., {Rickman}, E., {Sallum}, S., {Stapelfeldt}, K.~R., {Whiteford}, N., {Zhou}, Y., {Absil}, O., {Boccaletti}, A., {Booth}, M., {Bowler}, B.~P., {Chen}, C.~H., {Currie}, T., {Fortney}, J.~J., {Grady}, C.~A., {Greebaum}, A.~Z., {Henning}, T., {Hoch}, K. K.~W., {Janson}, M., {Kalas}, P., {Kenworthy}, M.~A., {Kervella}, P., {Kraus}, A.~L., {Lagage}, P.-O., {Liu}, M.~C., {Macintosh}, B., {Marino}, S., {Marley}, M.~S., {Marois}, C., {Matthews}, B.~C., {Mawet}, D., {McElwain}, M.~W., {Metchev}, S., {Meyer}, M.~R., {Molliere}, P., {Moran}, S.~E., {Morley}, C.~V., {Mukherjee}, S.,
  {Pantin}, E., {Quirrenbach}, A., {Rebollido}, I., {Ren}, B.~B., {Schneider}, G., {Vasist}, M., {Worthen}, K., {Wyatt}, M.~C., {Briesemeister}, Z.~W., {Bryan}, M.~L., {Calissendorff}, P., {Cantalloube}, F., {Cugno}, G., {De Furio}, M., {Dupuy}, T.~J., {Factor}, S.~M., {Faherty}, J.~K., {Fitzgerald}, M.~P., {Franson}, K., {Gonzales}, E.~C., {Hood}, C.~E., {Howe}, A.~R., {Kuzuhara}, M., {Lagrange}, A.-M., {Lawson}, K., {Lazzoni}, C., {Lew}, B. W.~P., {Liu}, P., {Llop-Sayson}, J., {Lloyd}, J.~P., {Martinez}, R.~A., {Mazoyer}, J., {Palma-Bifani}, P., {Quanz}, S.~P., {Redai}, J.~A., {Samland}, M., {Schlieder}, J.~E., {Tamura}, M., {Tan}, X., {Uyama}, T., {Vigan}, A., {Vos}, J.~M., {Wagner}, K., {Wolff}, S.~G., {Ygouf}, M., {Zhang}, X., {Zhang}, K., and {Zhang}, Z., ``{The JWST Early Release Science Program for Direct Observations of Exoplanetary Systems I: High-contrast Imaging of the Exoplanet HIP 65426 b from 2 to 16 {\ensuremath{\mu}}m},'' {\em \apjl}~{\bf 951},  L20 (July 2023).

\bibitem{GodoySPIE2024}
{Godoy}, N., {Choquet}, E., {Altinier}, L., {Lau}, A., {Mayer}, R., {Vigan}, A., and {Mary}, D., ``Escape project: fundamental detection limits of jwst/nircam coronographic observations,'' {\em \procspie} {\bf 13092} (2024).

\bibitem{Soummer2012}
{Soummer}, R., {Pueyo}, L., and {Larkin}, J., ``{Detection and Characterization of Exoplanets and Disks Using Projections on Karhunen-Lo{\`e}ve Eigenimages},'' {\em \apjl}~{\bf 755},  L28 (Aug. 2012).

\bibitem{Marois2006}
{Marois}, C., {Lafreni{\`e}re}, D., {Doyon}, R., {Macintosh}, B., and {Nadeau}, D., ``{Angular Differential Imaging: A Powerful High-Contrast Imaging Technique},'' {\em \apj}~{\bf 641},  556--564 (Apr. 2006).

\bibitem{Smith1984}
{Smith}, B.~A. and {Terrile}, R.~J., ``{A Circumstellar Disk around {\ensuremath{\beta}} Pictoris},'' {\em Science}~{\bf 226},  1421--1424 (Dec. 1984).

\bibitem{Potier2022}
{Potier}, A., {Mazoyer}, J., {Wahhaj}, Z., {Baudoz}, P., {Chauvin}, G., {Galicher}, R., and {Ruane}, G., ``{Increasing the raw contrast of VLT/SPHERE with the dark hole technique. II. On-sky wavefront correction and coherent differential imaging},'' {\em \aap}~{\bf 665},  A136 (Sept. 2022).

\bibitem{Bottom2017}
{Bottom}, M., {Wallace}, J.~K., {Bartos}, R.~D., {Shelton}, J.~C., and {Serabyn}, E., ``{Speckle suppression and companion detection using coherent differential imaging},'' {\em \mnras}~{\bf 464},  2937--2951 (Jan. 2017).

\bibitem{Baudoz2006}
{Baudoz}, P., {Boccaletti}, A., {Baudrand}, J., and {Rouan}, D., ``{The Self-Coherent Camera: a new tool for planet detection},'' in [{\em IAU Colloq. 200: Direct Imaging of Exoplanets: Science \& Techniques}{\nolinebreak\hspace{0.1em}]},  {Aime}, C. and {Vakili}, F., eds.,  553--558 (Jan. 2006).

\bibitem{Krist2023}
{Krist}, J.~E., {Steeves}, J.~B., {Dube}, B.~D., {Eldorado Riggs}, A.~J., {Kern}, B.~D., {Marx}, D.~S., {Cady}, E.~J., {Zhou}, H., {Poberezhskiy}, I.~Y., {Baker}, C.~W., {McGuire}, J.~P., {Nemati}, B., {Kuan}, G.~M., {Mennesson}, B., {Trauger}, J.~T., {Saini}, N.~S., and {Rafels}, S.~H., ``{End-to-end numerical modeling of the Roman Space Telescope coronagraph},'' {\em Journal of Astronomical Telescopes, Instruments, and Systems}~{\bf 9},  045002 (Oct. 2023).

\bibitem{Give'on2007SPIE}
{Give'on}, A., {Kern}, B., {Shaklan}, S., {Moody}, D.~C., and {Pueyo}, L., ``{Broadband wavefront correction algorithm for high-contrast imaging systems},'' in [{\em Astronomical Adaptive Optics Systems and Applications III}{\nolinebreak\hspace{0.1em}]},  {Tyson}, R.~K. and {Lloyd-Hart}, M., eds., {\em Society of Photo-Optical Instrumentation Engineers (SPIE) Conference Series} {\bf 6691},  66910A (Sept. 2007).

\bibitem{Giveon2011SPIE}
{Give'on}, A., {Kern}, B.~D., and {Shaklan}, S., ``{Pair-wise, deformable mirror, image plane-based diversity electric field estimation for high contrast coronagraphy},'' in [{\em Techniques and Instrumentation for Detection of Exoplanets V}{\nolinebreak\hspace{0.1em}]},  {Shaklan}, S., ed., {\em Society of Photo-Optical Instrumentation Engineers (SPIE) Conference Series} {\bf 8151},  815110 (Oct. 2011).

\bibitem{SoummerSPIE2022}
{Soummer}, R., {Por}, E.~H., {Pourcelot}, R., {Redmond}, S., {Laginja}, I., {Will}, S.~D., {Perrin}, M.~D., {Pueyo}, L., {Sahoo}, A., {Petrone}, P., {Brooks}, K.~J., {Fox}, R., {Klein}, A., {Nickson}, B., {Comeau}, T., {Ferrari}, M., {Gontrum}, R., {Hagopian}, J., {Leboulleux}, L., {Leongomez}, D., {Lugten}, J., {Mugnier}, L.~M., {N'Diaye}, M., {Nguyen}, M., {Noss}, J., {Sauvage}, J.-F., {Scott}, N., {Sivaramakrishnan}, A., {Subedi}, H.~B., and {Weinstock}, S., ``{High-contrast imager for complex aperture telescopes (HiCAT): 8. Dark zone demonstration with simultaneous closed-loop low-order wavefront sensing and control},'' in [{\em Space Telescopes and Instrumentation 2022: Optical, Infrared, and Millimeter Wave}{\nolinebreak\hspace{0.1em}]},  {Coyle}, L.~E., {Matsuura}, S., and {Perrin}, M.~D., eds., {\em Society of Photo-Optical Instrumentation Engineers (SPIE) Conference Series} {\bf 12180},  1218026 (Aug. 2022).

\bibitem{SoummerSPIE2024}
{Soummer}, R., {Pourcelot}, R., {Por}, E., {Steiger}, S., {Laginja}, I., {Buralli}, B., {Pueyo}, L., {Nguyen}, M., {Nickson}, B., {Sahoo}, A., and {the extended HiCAT team}, ``High-contrast imager for complex aperture telescopes (hicat): 11. system-level static and dynamic demonstration of the apodized pupil lyot coronagraph with a segmented aperture.,'' {\em \procspie} {\bf 13092} (2024).

\bibitem{catkit2}
Por, Emiel, H., Laginja, I., Pourcelot, R., Soummer, R., Sevin, A., Sahoo, A., Nguyen, M., Fowler, J., Egger, L., Pougheon, E., and Demagny, A., ``The control and automation for testbeds kit 2 (catkit2),'' (May 2024).

\bibitem{Por2018HighContrastImaging}
Por, E.~H., Haffert, S.~Y., Radhakrishnan, V.~M., Doelman, D.~S., van Kooten, M., and Bos, S.~P., ``{High Contrast Imaging for Python (HCIPy): an open-source adaptive optics and coronagraph simulator},'' in [{\em Adaptive Optics Systems VI}{\nolinebreak\hspace{0.1em}]},  Close, L.~M., Schreiber, L., and Schmidt, D., eds., {\em \procspie}~{\bf 10703, 1070342},  1112 -- 1125, International Society for Optics and Photonics, SPIE (2018).

\end{thebibliography}
\bibliographystyle{spiebib} 
\end{document}